\documentclass[aps,prl,showpacs,twocolumn,superscriptaddress]{revtex4}

\usepackage{graphicx}
\usepackage{mathptmx}

\begin{document}

\title{Spin-driven Phonon Splitting in Bond-frustrated ZnCr$_2$S$_4$}

\author{J. Hemberger}
\author{T. Rudolf}
\author{H.-A. Krug von Nidda}
\author{F.~Mayr}
\author{A. Pimenov}

\affiliation{Experimental Physics V, Center for Electronic
Correlations and Magnetism, University of Augsburg, D-86159
Augsburg, Germany}

\author{V. Tsurkan}

\thanks{Corresponding author}

\affiliation{Experimental Physics V, Center for Electronic
Correlations and Magnetism, University of Augsburg, D-86159
Augsburg, Germany}

\affiliation{Institute of Applied Physics, Academy of Sciences of
Moldova, MD-2028 Chi\c{s}in\v{a}u, R. Moldova}

\author{A. Loidl}

\affiliation{Experimental Physics V, Center for Electronic
Correlations and Magnetism, University of Augsburg, D-86159
Augsburg, Germany}

\begin{abstract}
Utilizing  magnetic susceptibility, specific heat,  thermal
expansion and IR spectroscopy we provide experimental evidence that
the  two subsequent antiferromagnetic transitions in ZnCr$_2$S$_4$
at $T_\mathrm{N1}$ = 15~K and $T_\mathrm{N2}$ = 8~K are accompanied
by significant thermal and phonon anomalies. The anomaly at
$T_\mathrm{N2}$ reveals a strong temperature hysteresis typical for
a first-order transformation. Due to strong spin-phonon coupling
both magnetic phase transitions induce a splitting of phonon modes,
where at $T_\mathrm{N1}$ the high-frequency and at $T_\mathrm{N2}$
the low-frequency modes split. The anomalies and phonon splitting
observed at $T_\mathrm{N2}$ are strongly suppressed by magnetic
field. Regarding the small positive Curie-Weiss temperature $\Theta
\simeq 8$~K, we argue that this scenario of two different magnetic
phases with concomitant different magneto-elastic couplings results
from the strong competition of ferromagnetic and antiferromagnetic
exchange of equal strength.
\end{abstract}


\maketitle

During the last decade the fascinating physics  of  spinel compounds
came into the focus of modern solid-state physics and materials
science. The classical and still unsolved nature  of the  Verwey
transition in magnetite, where charge and  orbital order compete
\cite{huang:04}, the formation of heavy fermions in LiV$_2$O$_4$
\cite{kondo:97}, the emergence of self-organized spin loops in
ZnCr$_2$O$_4$ \cite{lee:02}, the observation of colossal
magneto-resistance in Cu doped FeCr$_2$S$_4$ \cite{ramirez:97} and
of gigantic Kerr rotation in FeCr$_2$S$_4$ \cite{ogushi:05}, the
existence  of  an orbital glass state  in FeCr$_2$S$_4$
\cite{fichtl:05} and of a spin-orbital liquid  in FeSc$_2$S$_4$
\cite{fritsch:04}, multiferroic behavior and colossal
magneto-capacitive effect in CdCr$_2$S$_4$ and HgCr$_2$S$_4$
\cite{hemberger:05a}, spin dimerization in CuIr$_2$S$_4$
\cite{radaelli:02} and MgTi$_2$O$_4$ \cite{schmidt:04}, and the
spin-Peierls-like transitions  in 3-dimensional solids
\cite{lee:00,tchernyshyov:02a} are most representative examples of
exotic phenomena and ground states which were found recently in a
variety of spinel compounds. The appearance of these fascinating
ground states  is  attributed to the competition of charge, spin and
orbital degrees of freedom, which are strongly coupled to the
lattice. In addition, both A and B sites in the normal AB$_2$X$_4$
spinels are geometrically frustrated. Within the B-sublattice
further complexity is emerging due to the competition of nearest
neighbor (nn) ferromagnetic (FM) with direct as well as next-nearest
neighbor (nnn) antiferromagnetic (AFM) exchange. Hence, depending
only  on the B-site separation, FM and AFM ground states can  be
found \cite{menyuk:66,baltzer:66}, and in some cases FM and AFM
exchange interactions are  of equal strength leading  to  strong
frustration.

In this letter we present a detailed investigation of ZnCr$_2$S$_4$
comparing our results with those observed in other zinc-chromium
spinels. In these compounds Cr$^{3+}$ reveals a half-filled $t_{2g}$
crystal-field ground state with almost zero spin-orbit coupling.
Despite the fact that oxide, sulfide, and selenide are governed by
different exchange interactions, as indicated by their Curie-Weiss
(CW) temperatures, they reveal similar magnetic transition
temperatures into AFM states: ZnCr$_2$O$_4$, with the smallest Cr-Cr
separation, has a CW temperature of -390~K and exhibits a transition
from a paramagnet with strong quantum fluctuations into a complex
planar antiferromagnet at $T_\mathrm{N}$ = 12.5 K, accompanied by a
small tetragonal distortion \cite{lee:02,lee:00,chung:05}. The oxide
is governed by direct AFM Cr-Cr exchange \cite{goodenough:60}. The
ratio of CW to N\'{e}el temperature, defining the frustration
parameter f = $\Theta$ /$T_\mathrm{N} \sim 30$, signals strong
geometrical frustration \cite{ramirez:01}. The structural phase
transition has been explained in terms of a spin Jahn-Teller effect
\cite{lee:00,tchernyshyov:02a, yamashita:00}. The selenide,
ZnCr$_2$Se$_4$, with the largest Cr-Cr distance of the zinc-chromium
spinels, has a positive CW temperature of 115 K and exhibits a
transition into a helical magnetic structure at 20 K
\cite{menyuk:66}. The helical structure is characterized by
ferromagnetic (001) planes with a propagation vector along the [001]
axis \cite{plumier:66}.  The increase of the lattice constant in the
selenide almost suppresses the direct exchange and the spin
arrangement follows from the dominating ferromagnetic 90$^\circ$
Cr-Se-Cr exchange and the additional influence of AFM interactions
(some reminder of the direct exchange, plus nnn Cr-S-Zn-S-Cr and
Cr-S-S-Cr exchange). Again, the magnetic phase transition is
accompanied by a small tetragonal distortion with $1-c/a = 0.001$
\cite{kleinberger:66}.

The sulfide, ZnCr$_2$S$_4$, lies exactly in between these two
extremes. Consequently, FM and AFM exchange interactions almost
compensate each other yielding a CW temperature of approximately 0
K. Neutron-scattering experiments \cite{hamedoun:86a} reveal a
magnetic phase transition at 15.5~K into a helical spin order very
similar to that in the selenide. Below 12~K a second commensurate
collinear antiferromagnetic phase starts to gradually develop down
to 8~K. This second phase has a similar spin arrangement like the
AFM oxide. At low temperatures both magnetic phases coexist
\cite{hamedoun:86a}. This fact already documents that both spin
structures are characterized by almost the same free energy. The
metastable magnetic ground state comes along with the competing FM
and AFM exchange of almost equal strength, a situation we term as
bond frustration. So far structural phase transitions in
ZnCr$_2$S$_4$ have not been reported to occur along with the
magnetic transitions. In reference \cite{hamedoun:86a} an upper
limit of $1-c/a < 0.002$ has been established. However, we would
like to note that in all Zn-Cr spinels the structural phase
transition yields marginal distortions only and can hardly be
detected using standard diffraction techniques. Here we use infrared
spectroscopy to probe structural transitions and spin-lattice
correlations.

Polycrystalline ZnCr$_2$S$_4$ was prepared by solid-state reaction
from high purity elements at 800$^\circ$C. X-ray diffraction
analysis at room temperature revealed single-phase material with the
cubic  spinel structure with a lattice constant $a = 9.983(2)$~{\AA}
and a sulfur fractional coordinate $x = 0.258(1)$. The magnetic
properties were measured using a commercial SQUID magnetometer
(Quantum Design MPMS-5). The heat capacity was monitored in a
Quantum Design PPMS for temperatures $2~\mathrm{K} < T <
300~\mathrm{K}$ and in external magnetic fields up to 70~kOe.  The
thermal expansion was measured by capacitive method in fields up to
70~kOe. The reflectivity experiments on ceramic samples with
polished surfaces were carried out in the far infrared (FIR) range
using the Fourier-transform spectrometer Bruker IFS~113v in a He
bath cryostat.

\begin{figure}
\centering
 \includegraphics[width=0.40\textwidth,clip]{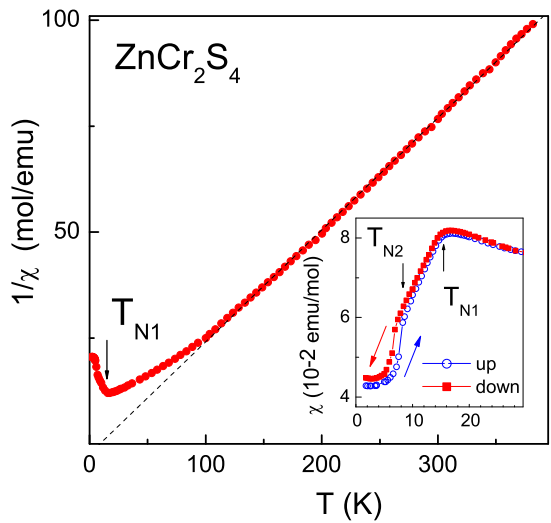}
 \caption{(color online) Inverse susceptibility vs temperature as measured in
 ZnCr$_2$S$_4$ at 10~kOe. The dashed line indicates a Curie-Weiss behavior.
  Inset: Low-temperature susceptibility for heating (up) and cooling (down)
 cycles. Arrows indicate the magnetic phase transitions at $T_\mathrm{N1}$ and  $T_\mathrm{N2}$.}
  \label{fig1}
\end{figure}

Figure~\ref{fig1} presents the inverse susceptibility $\chi^{-1}$ vs
temperature. For $T > 100~\mathrm{K}$, $\chi$ follows a Curie-Weiss
law with a positive CW temperature $\Theta \simeq 8$~K and an
effective moment of 3.86~$\mu_\mathrm{B}$, in good agreement with
the spin-only value of Cr$^{3+}$ in $3d^{3}$ configuration. In
preliminary ESR experiments we determined a $g$-value of 1.98,
indeed indicating negligible spin-orbit coupling. Below 100~K,
$\chi^{-1}(T)$ deviates from the FM CW law suggesting an increasing
contribution of the AFM spin correlations. At 15~K the inverse
susceptibility shows a clear minimum. This temperature marks the
onset of long-range antiferromagnetic order at $T_\mathrm{N1}$, as
found by earlier studies \cite{hamedoun:86a}.

The inset in Fig.~\ref{fig1} provides a closer look to the
transition region. In a narrow temperature range around
$T_\mathrm{N1}$, our results are similar to those obtained on
polycrystalline and single-crystalline samples in
\cite{hamedoun:86a}. However, below $T_\mathrm{N1}$  after a
continuous drop the susceptibility exhibits a sharp change of slope
at around $T_\mathrm{N2}$ = 8~K. In addition, we observed a
pronounced difference between the susceptibilities measured on
cooling and heating being maximal at $T_\mathrm{N2}$. Such a
hysteresis indicates a first-order transformation. Both magnetic
anomalies are in good agreement with the early neutron-diffraction
experiments \cite{hamedoun:86a}. In the following we demonstrate
that the anomalies in the susceptibility  correlate with respective
anomalies found in the specific heat, thermal expansion and IR
spectra.

\begin{figure}
\centering
 \includegraphics[width=0.40\textwidth,clip]{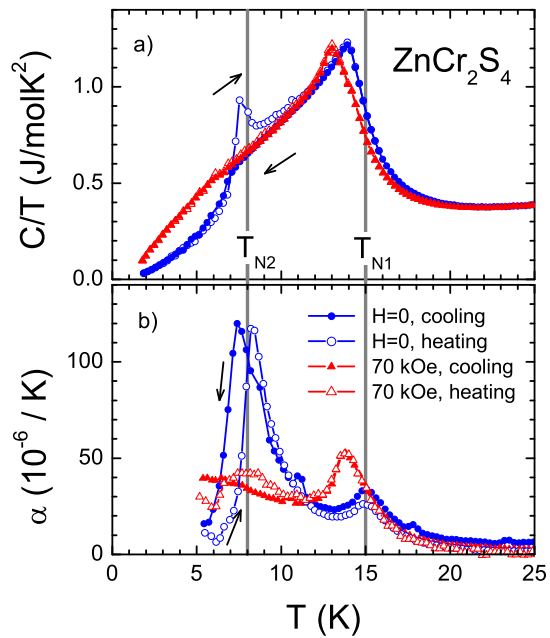}
 \caption{(color online) a) Heat capacity of ZnCr$_2$S$_4$ plotted as $C/T$ vs $T$
 measured on cooling (closed symbols) and heating (open symbols) for 0 and 70~kOe.
  b) Temperature dependence  of the  thermal expansion coefficient
 at different external magnetic fields. Vertical lines indicate the phase transitions
 $T_\mathrm{N1}$ and  $T_\mathrm{N2}$ in zero field.}
 \label{fig2}
\end{figure}

Figure~\ref{fig2}a shows the specific heat in the representation
$C/T$ vs temperature at different external magnetic fields. On
approaching the N\'{e}el temperature $T_\mathrm{N1}$, the specific
heat manifests  a sharp $\lambda$-type anomaly. At lower
temperature, a second anomaly at $T_\mathrm{N2}$ becomes evident as
a peak on heating and a kink on cooling. Note that this is the
ordinary way in which the hysteresis at a first-order transition
becomes evident in the relaxation method used \cite{hemberger:05b}.
The anomaly at $T_\mathrm{N2}$ in the specific heat correlates with
the respective anomaly observed in the susceptibility. Strong
spin-lattice coupling and possible structural instabilities
connected with both magnetic transitions can be deduced from the
thermal expansion which exhibits significant anomalies both at
$T_\mathrm{N1}$ and $T_\mathrm{N2}$ (Fig.~\ref{fig2}b). In zero
magnetic field the thermal expansion coefficient $\alpha$ at the
lower transition is by a factor of 4 larger and again reveals a
significant hysteresis in accordance with the first-order character
and, thus, the importance of the spin-lattice correlations. Magnetic
field has a strong effect on the specific heat and thermal
expansion, shifting the maximum in $C/T$ at $T_\mathrm{N1}$ to lower
temperatures as usually observed in antiferromagnets. A shift to
lower $T$ is found also for the maximum in the thermal expansion
coefficient  at $T_\mathrm{N1}$ which in addition shows an increase
in the peak height. At the same time, at $T_\mathrm{N2}$ the peaks
in $C/T$ and $\alpha$ become strongly suppressed above fields of
20~kOe. Assuming that the peak at $T_\mathrm{N1}$ is related to the
onset of a helical structure, the field dependence of the peak in
the thermal expansion (Fig. 2b) suggests a stabilization of the
helical spin arrangement in the field. We note that the overall
magnetic structure is AFM, however the magnetic field favors the
helical configuration due to the induced FM component. This
corroborates our interpretation that the helical spin structure in
ZnCr$_2$S$_4$ is supported by FM exchange, as already documented in
ZnCr$_2$Se$_4$, which has a large and positive CW temperature and
reveals a helical spin structure as stable ground state. At the same
time, the complex collinear structure below $T_\mathrm{N2}$ is
supported by AFM exchange, in close analogy to strongly
geometrically frustrated ZnCr$_2$O$_4$. In ZnCr$_2$S$_4$, the
external magnetic fields break the balance of FM and AFM exchange
interactions in favor of FM exchange and, hence, stabilizes the
helical spin arrangement as revealed by the specific-heat and
thermal-expansion experiments. The existence of strong spin-phonon
coupling, which follows from the thermal-expansion data, is
microscopically probed by means of IR spectroscopy.

\begin{figure}
\centering
 \includegraphics[width=0.47\textwidth,clip]{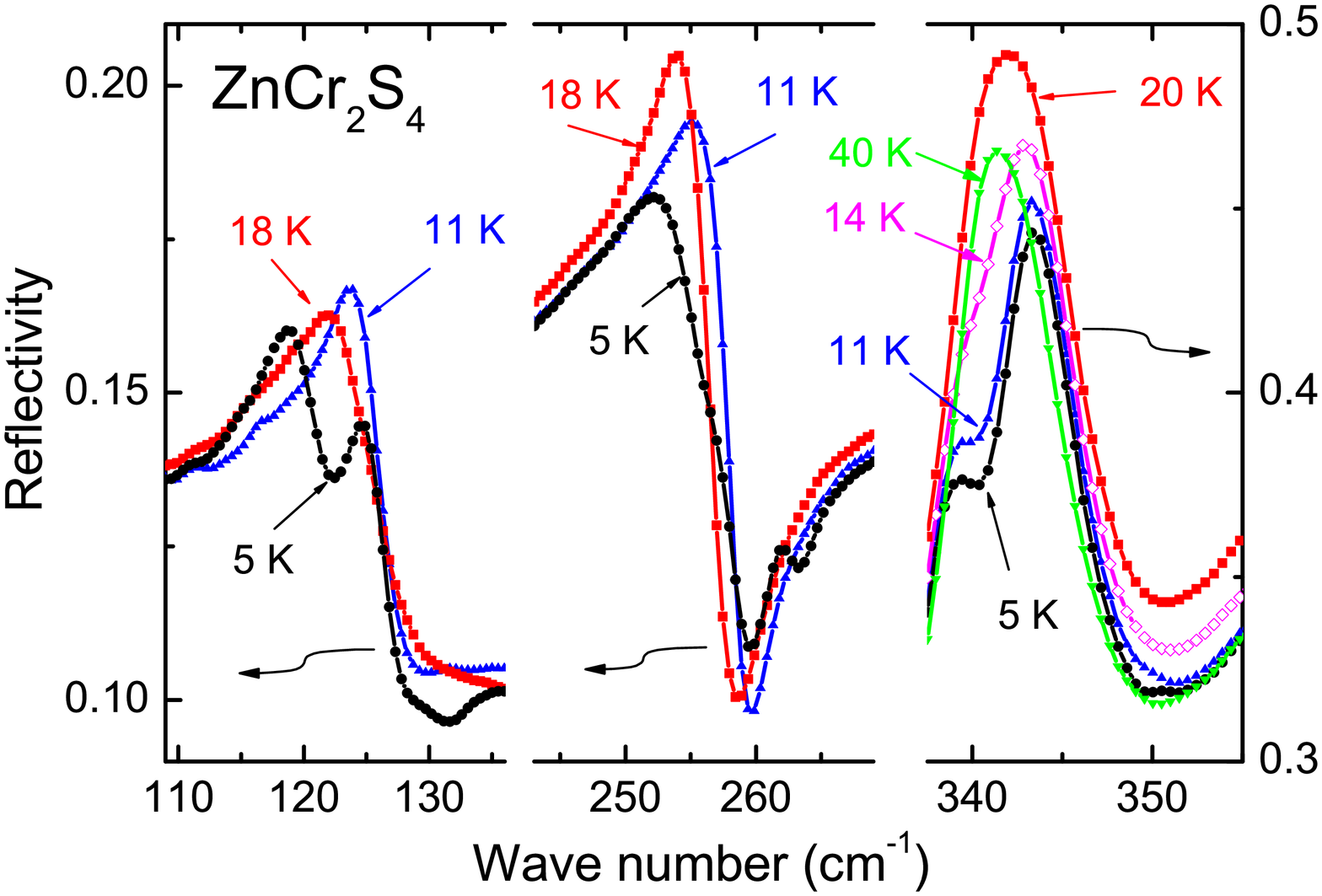}
 \caption{(color online) Reflectivity for three phonon modes of
  ZnCr$_2$S$_4$ at temperatures around the magnetic transitions.}
 \label{fig3}
\end{figure}

For spinel compounds which show strong spin-phonon coupling effects
\cite{baltensperger:69,bruesch:72,wakamura:88,wakamura:89} it is
well established that the sign of shift of the phonon
eigenfrequencies in the magnetically  ordered state primarily
depends on the type  of coupling: ferromagnetic  or
antiferromagnetic, with the latter revealing positive shifts of the
eigenfrequencies \cite{wakamura:88,wakamura:89}. The main magnetic
coupling mechanism in sulfides is superexchange  which includes nn
FM Cr-S-Cr  and nnn AFM Cr-S-Zn-S-Cr or Cr-S-S-Cr exchange
interactions \cite{menyuk:66,baltzer:66}. From a thorough eigenmode
analysis \cite{bruesch:72} it is known that the low-lying modes
involve vibration of the Zn-S units, while Cr-S atoms are mostly
involved in the high-frequency modes.

For the spinels with FM ground states like CdCr$_2$S$_4$
\cite{wakamura:88} as well as for ferrimagnetic FeCr$_2$S$_4$
\cite{wakamura:89}, it has been demonstrated that the low lying
modes reveal large and positive shifts due to AFM exchange, while
negative shifts have been observed for the high-frequency modes. But
in all these cases no splitting of the phonon modes could be
detected at $T_\mathrm{c}$, which seems to be clear as the FM spin
order does not involve any lattice-symmetry breaking. A significant
phonon splitting has been reported at the AFM phase transition in
ZnCr$_2$O$_4$ which is accompanied by a small tetragonal distortion
\cite{sushkov:05}. Recently, this phonon splitting has been
interpreted as driven only by spin correlations in the collinear AFM
spin structure, where the magnetically induced symmetry breaking
appears only in the dynamic phonon properties, but preserving the
cubic structure \cite{fennie:06}. The effect of phonon splitting due
to magnetic ordering even in the absence of any structural symmetry
breaking has been predicted earlier by Massidda \textit{et al.}
\cite{massidda:99}.

Our room-temperature IR results on ZnCr$_2$S$_4$ are in good
agreement with previously published data \cite{lutz:83}. The IR
spectrum reveals the four group-theoretically allowed phonon modes.
A fit using  a sum of Lorentz oscillators yields eigenfrequencies at
115, 244, 336, and 388 ~cm$^{-1}$. On decreasing temperature we find
a positive frequency shift of the phonon eigenmodes, but below the
magnetic phase  transition temperatures all phonon modes reveal a
clear splitting indicating a symmetry breaking as illustrated in
Fig. 3. The eigenfrequency for all phonon modes as function of
temperature is presented in Fig. 4 on a semilogarithmic plot. From
room temperature down to 20 K the purely anharmonic behavior of a
cubic spinel naturally describes the results. Below the first
magnetic transition into the helical structure all main modes reveal
an additional shift towards higher frequencies, as expected for AFM
transitions.  The shift is of the order of 0.5\% for the high and of
the order of 1\% for the low-frequency modes.  As the most important
result we find that the splitting of the phonon modes does not occur
at the same temperature: The high-frequency modes reveal significant
splitting certainly at $T_\mathrm{N1}$ = 15 K: A weak but clearly
pronounced shoulder is visible already at 14 K. At 5 K the splitting
of both high-frequency phonons amounts approximately 1\%. Contrary,
the two low-frequency modes split distinctly below $T_\mathrm{N1}$.
The modes at 125 and 255 ~cm$^{-1}$ reveal single-mode behavior
still at 11 K, and split at $T$ close to $T_\mathrm{N2}$ = 8 K. At 5
K the mode splitting results in extra peaks for both modes and in a
significant shoulder for the phonon at 260~cm$^{-1}$, which
indicates a splitting into three modes at low temperatures.  The
overall splitting is of the order of 5\% for both low-frequency
modes.

The splitting of the IR-active modes in ZnCr$_2$S$_4$ at the two
magnetic phase transitions further supports our interpretation. As
outlined above, the high-frequency modes involve vibrations mostly
of Cr-S atoms. The FM 90$^\circ$ Cr-S-Cr exchange governs the
helical order of ferromagnetic (001) planes. This exchange
interaction is responsible for the splitting of the high-frequency
modes. But ZnCr$_2$S$_4$ is equally dominated by AFM exchange which
establishes the complex commensurate collinear spin order. These AFM
exchange interactions involve mainly Zn-S bonds which on the other
side are involved in eigenfrequencies of the low-lying modes:
Consequently, at $T_\mathrm{N2}$ these modes are expected to split
as we experimentally observed. The comparable splitting of both
high- and low-frequency modes ($\sim5$~cm$^{-1}$) provides
compelling evidence that FM and AFM exchange interactions are of
equal strength. The ground state energies of the two spin
configurations are similar and the transition into the low
temperature phase is generated by these competing interactions.
Preliminary experiments in external magnetic fields show that the
splitting of the low-frequency phonon modes is strongly suppressed
for fields above 50 kOe. This observation is in agreement with the
reduction of the anomalies of the specific heat and thermal
expansion at $T_\mathrm{N2}$ by an external magnetic field. Thus,
our experiments prove that the low-temperature symmetry breaking in
the Zn-Cr spinels mainly results from a strong spin-phonon coupling.
This has been established for the geometrically frustrated
ZnCr$_2$O$_4$ and we demonstrate this for the bond frustrated
ZnCr$_2$S$_4$. It will be highly interesting to see, if the phonon
splitting is accompanied by two consecutive structural
transformations into low-symmetry structural phases or the symmetry
breaking can only be observed via dynamic variables. High-resolution
synchrotron experiments on single crystals will probably be
necessary to solve this question.

\begin{figure}
\centering
 \includegraphics[width=0.38\textwidth,clip]{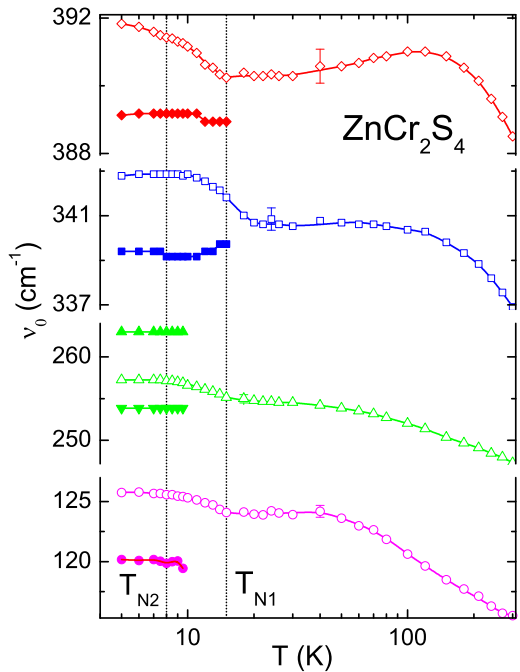}
 \caption{(color online) Temperature dependence of the eigenfrequencies
 of the all phonon modes on a semilogarithmic plot.
 Open symbols represent the mode of the
 high-temperature phase; closed symbols - the new modes which appear at $T_\mathrm{N1}$ and $T_\mathrm{N2}$.}
 \label{fig4}
\end{figure}

In conclusion, we investigated the bond frustrated AFM spinel
ZnCr$_2$S$_4$ and found a splitting of high-frequency phonons below
the onset of the antiferromagnetic helical order and splitting of
the low-frequency modes at around the second magnetic transition
into the collinear commensurate phase. Magnetic susceptibility,
specific heat and thermal expansion show an additional anomaly at
the second transition beside the anomaly at the N\'{e}el point. Our
results reveal strong spin-phonon coupling that generates the
low-temperature instability in ZnCr$_2$S$_4$. Suppression of the
splitting of low-frequency phonons and concomitantly of the
anomalies in the specific heat and thermal expansion by strong
magnetic fields suggests a spin-driven origin of this
transformation.

%\begin{acknowlegments}
We are grateful to Dana Vieweg, Veronika Fritsch,  Natalia Tristan,
and Thomas Wiedenmann for experimental support. This work was
supported by BMBF via VDI/EKM, FKZ 13N6917-B and by DFG within SFB
484 (Augsburg).
%\end{acknowlegments}


\begin{thebibliography}{39}

\expandafter\ifx\csname natexlab\endcsname\relax\def\natexlab#1{#1}\fi
\expandafter\ifx\csname bibnamefont\endcsname\relax
  \def\bibnamefont#1{#1}\fi
\expandafter\ifx\csname bibfnamefont\endcsname\relax
  \def\bibfnamefont#1{#1}\fi
\expandafter\ifx\csname citenamefont\endcsname\relax
  \def\citenamefont#1{#1}\fi
\expandafter\ifx\csname url\endcsname\relax
  \def\url#1{\texttt{#1}}\fi
\expandafter\ifx\csname urlprefix\endcsname\relax\def\urlprefix{URL }\fi
\providecommand{\bibinfo}[2]{#2} \providecommand{\eprint}[2][]{\url{#2}}


\bibitem[{\citenamefont{Huang et~al.}(2004)\citenamefont{Huang et~al.}}]{huang:04}
  \bibinfo{author}{\bibnamefont{D. J.}~\bibnamefont{Huang}} \emph{et~al.},
  \bibinfo{journal}{Phys. Rev. Lett.} \textbf{\bibinfo{volume}{93}},
  \bibinfo{pages}{077204} (\bibinfo{year}{2004}); \bibinfo{author}{\bibnamefont{I.}~\bibnamefont{Leonov}} \emph{et~al.},
  \bibinfo{journal}{Phys. Rev. Lett.} \textbf{\bibinfo{volume}{93}},
  \bibinfo{pages}{146404} (\bibinfo{year}{2004}).

\bibitem[{\citenamefont{Kondo et~al.}(1997)\citenamefont{Kondo et~al.}}]{kondo:97}
  \bibinfo{author}{\bibnamefont{S.}~\bibnamefont{Kondo}} \emph{et~al.},
  \bibinfo{journal}{Phys. Rev. Lett.} \textbf{\bibinfo{volume}{78}},
  \bibinfo{pages}{3729} (\bibinfo{year}{1997}); \bibinfo{author}{\bibnamefont{A.}~\bibnamefont{Krimmel}} \emph{et~al.},
  \bibinfo{journal}{Phys. Rev. Lett.} \textbf{\bibinfo{volume}{82}},
  \bibinfo{pages}{2919} (\bibinfo{year}{1999}).

\bibitem[{\citenamefont{Lee et~al.}(2002)\citenamefont{Lee et~al.}}]{lee:02}
  \bibinfo{author}{\bibnamefont{S.-H.}~\bibnamefont{Lee}} \emph{et~al.},
  \bibinfo{journal}{Nature (London)} \textbf{\bibinfo{volume}{418}},
  \bibinfo{pages}{856} (\bibinfo{year}{2002}).

\bibitem[{\citenamefont{Ramirez et~al.}(1997)\citenamefont{Ramirez, Cava, Krajewski}}]{ramirez:97}
  \bibinfo{author}{\bibnamefont{A.~P.}~\bibnamefont{Ramirez}},
  \bibinfo{author}{\bibnamefont{R.~J.}~\bibnamefont{Cava}}, \bibnamefont{and}
  \bibinfo{author}{\bibnamefont{J.}~\bibnamefont{Krajewski}},
  \bibinfo{journal}{Nature (London)} \textbf{\bibinfo{volume}{386}},
  \bibinfo{pages}{156} (\bibinfo{year}{1997}); \bibinfo{author}{\bibnamefont{V.}~\bibnamefont{Fritsch}} \emph{et~al.},
  \bibinfo{journal}{Phys. Rev.~B} \textbf{\bibinfo{volume}{67}},
  \bibinfo{pages}{144419} (\bibinfo{year}{2003}).

\bibitem[{\citenamefont{Ogushi et~al.}(2005)\citenamefont{Ogushi et~al.}}]{ogushi:05}
  \bibinfo{author}{\bibnamefont{K.}~\bibnamefont{Ogushi}} \emph{et~al.},
  \bibinfo{journal}{Phys. Rev.~B} \textbf{\bibinfo{volume}{72}},
  \bibinfo{pages}{155114} (\bibinfo{year}{2005}).

\bibitem[{\citenamefont{Fichtl et~al.}(2005)\citenamefont{Fichtl et~al.}}]{fichtl:05}
  \bibinfo{author}{\bibnamefont{R.}~\bibnamefont{Fichtl}} \emph{et~al.},
  \bibinfo{journal}{Phys. Rev. Lett.} \textbf{\bibinfo{volume}{94}},
  \bibinfo{pages}{027601} (\bibinfo{year}{2005}).

\bibitem[{\citenamefont{Fritsch et~al.}(2004)\citenamefont{Fritsch et~al.}}]{fritsch:04}
  \bibinfo{author}{\bibnamefont{V.}~\bibnamefont{Fritsch}} \emph{et~al.},
  \bibinfo{journal}{Phys. Rev. Lett.} \textbf{\bibinfo{volume}{92}},
  \bibinfo{pages}{116401} (\bibinfo{year}{2004}); \bibinfo{author}{\bibnamefont{A.}~\bibnamefont{Krimmel}} \emph{et~al.},
  \bibinfo{journal}{Phys. Rev. Lett.} \textbf{\bibinfo{volume}{94}},
  \bibinfo{pages}{237402} (\bibinfo{year}{2005}).

\bibitem[{\citenamefont{Hemberger et~al.}(2005)\citenamefont{Hemberger et~al.}}]{hemberger:05a}
  \bibinfo{author}{\bibnamefont{J.}~\bibnamefont{Hemberger}} \emph{et~al.},
  \bibinfo{journal}{Nature (London)} \textbf{\bibinfo{volume}{416}},
  \bibinfo{pages}{364} (\bibinfo{year}{2005}); \bibinfo{author}{\bibnamefont{S.}~\bibnamefont{Weber}} \emph{et~al.},
  \bibinfo{journal}{Phys. Rev. Lett.} \textbf{\bibinfo{volume}{96}},
  \bibinfo{pages}{157202} (\bibinfo{year}{2006}).


\bibitem[{\citenamefont{Radaelli et~al.}(2002)\citenamefont{Radaelli et~al.}}]{radaelli:02}
  \bibinfo{author}{\bibnamefont{P.~G.}~\bibnamefont{Radaelli}} \emph{et~al.},
  \bibinfo{journal}{Nature (London)} \textbf{\bibinfo{volume}{416}},
  \bibinfo{pages}{155} (\bibinfo{year}{2002}).

\bibitem[{\citenamefont{Schmidt et~al.}(2004)\citenamefont{Schmidt et~al.}}]{schmidt:04}
  \bibinfo{author}{\bibnamefont{M.}~\bibnamefont{Schmidt}} \emph{et~al.},
  \bibinfo{journal}{Phys. Rev. Lett.} \textbf{\bibinfo{volume}{92}},
  \bibinfo{pages}{056402} (\bibinfo{year}{2004}).


\bibitem[{\citenamefont{Lee et~al.}(2000)\citenamefont{Lee et~al.}}]{lee:00}
  \bibinfo{author}{\bibnamefont{S.-H.}~\bibnamefont{Lee}} \emph{et~al.},
  \bibinfo{journal}{Phys. Rev. Lett.} \textbf{\bibinfo{volume}{84}},
  \bibinfo{pages}{3718} (\bibinfo{year}{2000}).


\bibitem[{\citenamefont{Tchernyshyov et~al.}(2002)\citenamefont{Tchernyshyov et~al.}}]{tchernyshyov:02a}
  \bibinfo{author}{\bibnamefont{O.}~\bibnamefont{Tchernyshyov}} \emph{et~al.},
  \bibinfo{journal}{Phys. Rev. Lett.} \textbf{\bibinfo{volume}{88}},
  \bibinfo{pages}{067203} (\bibinfo{year}{2002}); \bibinfo{author}{\bibnamefont{O.}~\bibnamefont{Tchernyshyov}} \emph{et~al.},
  \bibinfo{journal}{Phys. Rev.~B} \textbf{\bibinfo{volume}{66}},
  \bibinfo{pages}{064403} (\bibinfo{year}{2002}).

\bibitem[{\citenamefont{Menyuk et~al.}(1966)\citenamefont{Menyuk et~al.}}]{menyuk:66}
  \bibinfo{author}{\bibnamefont{N.}~\bibnamefont{Menyuk}} \emph{et~al.},
  \bibinfo{journal}{J. Appl. Phys.} \textbf{\bibinfo{volume}{37}},
  \bibinfo{pages}{1387} (\bibinfo{year}{1966}).

\bibitem[{\citenamefont{Baltzer et~al.}(1966)\citenamefont{Baltzer et~al.}}]{baltzer:66}
  \bibinfo{author}{\bibnamefont{K.}~\bibnamefont{Baltzer}} \emph{et~al.},
  \bibinfo{journal}{Phys. Rev.} \textbf{\bibinfo{volume}{151}},
  \bibinfo{pages}{367} (\bibinfo{year}{1966}).

\bibitem[{\citenamefont{Chung et~al.}(2005)\citenamefont{Chung et~al.}}]{chung:05}
  \bibinfo{author}{\bibnamefont{J.-H.}~\bibnamefont{Chung}} \emph{et~al.},
  \bibinfo{journal}{Phys. Rev. Lett.} \textbf{\bibinfo{volume}{95}},
  \bibinfo{pages}{247204} (\bibinfo{year}{2005}).


\bibitem[{\citenamefont{Googenough}(1960)\citenamefont{Googenough}}]{goodenough:60}
  \bibinfo{author}{\bibnamefont{J. B.}~\bibnamefont{Googenough}} ,
  \bibinfo{journal}{Phys. Rev. ~B} \textbf{\bibinfo{volume}{117}},
  \bibinfo{pages}{1442} (\bibinfo{year}{1960}).

\bibitem[{\citenamefont{Ramirez}(2001)}]{ramirez:01}
\bibinfo{author}{\bibfnamefont{A.~P.} \bibnamefont{Ramirez}}, in
  \emph{\bibinfo{booktitle}{Handbook of Magnetic Materials}}, edited by
  \bibinfo{editor}{\bibfnamefont{K.~H.~J.}~\bibnamefont{Buschow}}
  (\bibinfo{publisher}{Elsevier Science}, \bibinfo{address}{Amsterdam},
  \bibinfo{year}{2001}), Vol. \bibinfo{volume}{13}, p.~\bibinfo{pages}{423}.

\bibitem[{\citenamefont{Yamashita et~al.}(200)\citenamefont{Yamashita and Arai}}]{yamashita:00}
  \bibinfo{author}{\bibnamefont{Y.}~\bibnamefont{Yamashita}} \bibnamefont{and}
  \bibinfo{author}{\bibnamefont{K.}~\bibnamefont{Ueda}},
  \bibinfo{journal}{Phys. Rev. Lett.} \textbf{\bibinfo{volume}{85}},
  \bibinfo{pages}{4960} (\bibinfo{year}{2000}).

\bibitem[{\citenamefont{Plumier}(1966)\citenamefont{Plumier}}]{plumier:66}
  \bibinfo{author}{\bibnamefont{R.}~\bibnamefont{Plumier}},
  \bibinfo{journal}{J. Physique} \textbf{\bibinfo{volume}{27}},
  \bibinfo{pages}{213} (\bibinfo{year}{1966}).

\bibitem[{\citenamefont{Kleinberger et~al.}(1966)\citenamefont{Kleinberger and de Kouchkovsky}}]{kleinberger:66}
  \bibinfo{author}{\bibnamefont{R.}~\bibnamefont{Kleinberger}} \bibnamefont{and}
  \bibinfo{author}{\bibnamefont{R.}~\bibnamefont{de Kouchkovsky}},
  \bibinfo{journal}{C.R. Acad. Sci. Paris} \textbf{\bibinfo{volume}{262}},
  \bibinfo{pages}{628} (\bibinfo{year}{1966}); \bibinfo{author}{\bibnamefont{M.}~\bibnamefont{Hidaka}} \emph{et~al.},
  \bibinfo{journal}{phys. stat. sol. (b)} \textbf{\bibinfo{volume}{236}},
  \bibinfo{pages}{570} (\bibinfo{year}{2003}).

\bibitem[{\citenamefont{Hamedoun et~al.}(1986)\citenamefont{Hamedoun et~al.}}]{hamedoun:86a}
  \bibinfo{author}{\bibnamefont{M.}~\bibnamefont{Hamedoun}} \emph{et~al.},
  \bibinfo{journal}{J. Phys.~C} \textbf{\bibinfo{volume}{19}},
  \bibinfo{pages}{1783} (\bibinfo{year}{1986}); \bibinfo{author}{\bibnamefont{M.}~\bibnamefont{Hamedoun}} \emph{et~al.},
  \bibinfo{journal}{J. Phys.~C} \textbf{\bibinfo{volume}{19}},
  \bibinfo{pages}{1801} (\bibinfo{year}{1986}).

\bibitem[{\citenamefont{Hemberger et~al.}(2005)\citenamefont{Hemberger et~al.}}]{hemberger:05b}
  \bibinfo{author}{\bibnamefont{J.}~\bibnamefont{Hemberger}} \emph{et~al.},
  \bibinfo{journal}{Phys. Rev.~B} \textbf{\bibinfo{volume}{72}},
  \bibinfo{pages}{012420} (\bibinfo{year}{2005}); \bibinfo{author}{\bibnamefont{J.C.}~\bibnamefont{Lashley}} \emph{et~al.},
  \bibinfo{journal}{Cryogenics} \textbf{\bibinfo{volume}{43}},
  \bibinfo{pages}{369} (\bibinfo{year}{2003}).

\bibitem[{\citenamefont{Baltensperger et~al.}(1969)\citenamefont{Baltensperger and Helman}}]{baltensperger:69}
  \bibinfo{author}{\bibnamefont{W.}~\bibnamefont{Baltensperger}} \bibnamefont{and}
  \bibinfo{author}{\bibnamefont{J.~S.}~\bibnamefont{Helman}},
  \bibinfo{journal}{Helv. Phys. Acta} \textbf{\bibinfo{volume}{42}},
  \bibinfo{pages}{611} (\bibinfo{year}{1969}); \bibinfo{author}{\bibnamefont{W.}~\bibnamefont{Baltensperger}},
  \bibinfo{journal}{J. Appl. Phys.} \textbf{\bibinfo{volume}{41}},
  \bibinfo{pages}{1052} (\bibinfo{year}{1970}).

\bibitem[{\citenamefont{Bruesch et~al.}(1972)\citenamefont{Bruesch and D'Ambrogio}}]{bruesch:72}
  \bibinfo{author}{\bibnamefont{P.}~\bibnamefont{Bruesch}} \bibnamefont{and}
  \bibinfo{author}{\bibnamefont{F.}~\bibnamefont{D'Ambrogio}},
  \bibinfo{journal}{Phys. Stat. Sol. B} \textbf{\bibinfo{volume}{50}},
  \bibinfo{pages}{513} (\bibinfo{year}{1972}).

\bibitem[{\citenamefont{Wakamura et~al.}(1988)\citenamefont{Wakamura and Arai}}]{wakamura:88}
  \bibinfo{author}{\bibnamefont{K.}~\bibnamefont{Wakamura}} \bibnamefont{and}
  \bibinfo{author}{\bibnamefont{T.}~\bibnamefont{Arai}},
  \bibinfo{journal}{J. Appl. Phys.} \textbf{\bibinfo{volume}{63}},
  \bibinfo{pages}{5824} (\bibinfo{year}{1988}).

\bibitem[{\citenamefont{Wakamura}(1989)\citenamefont{Wakamura}}]{wakamura:89}
  \bibinfo{author}{\bibnamefont{K.}~\bibnamefont{Wakamura}}
  \bibinfo{journal}{Solid State Commun.} \textbf{\bibinfo{volume}{71}},
  \bibinfo{pages}{1033} (\bibinfo{year}{1989}); \bibinfo{author}{\bibnamefont{T.}~\bibnamefont{Rudolf}} \emph{et~al.},
  \bibinfo{journal}{Phys. Rev.~B} \textbf{\bibinfo{volume}{72}},
  \bibinfo{pages}{014450} (\bibinfo{year}{2005}).


\bibitem[{\citenamefont{Sushkov et~al.}(2005)\citenamefont{Sushkov et~al.}}]{sushkov:05}
  \bibinfo{author}{\bibnamefont{A.~B.}~\bibnamefont{Sushkov}} \emph{et~al.},
  \bibinfo{journal}{Phys. Rev. Lett.} \textbf{\bibinfo{volume}{94}},
  \bibinfo{pages}{137202} (\bibinfo{year}{2005}).

\bibitem[{\citenamefont{Fennie et~al.}(1988)\citenamefont{Fennie and Rabe}}]{fennie:06}
  \bibinfo{author}{\bibnamefont{C. J.}~\bibnamefont{Fennie}} \bibnamefont{and}
  \bibinfo{author}{\bibnamefont{K.}~\bibnamefont{Rabe}},
  \bibinfo{journal}{cond-mat}/%\textbf{\bibinfo{volume}{1}},
  \bibinfo{pages}{0602503}. %(\bibinfo{year}{2006}).

\bibitem[{\citenamefont{Massidda et~al.}(200)\citenamefont{Massidda et~al.}}]{massidda:99}
  \bibinfo{author}{\bibnamefont{S.}~\bibnamefont{Massidda}} \emph{et~al.},
  \bibinfo{journal}{Phys. Rev. Lett.} \textbf{\bibinfo{volume}{82}},
  \bibinfo{pages}{430} (\bibinfo{year}{1999}).


\bibitem[{\citenamefont{Lutz et~al.}(1983)\citenamefont{Lutz et~al.}}]{lutz:83}
  \bibinfo{author}{\bibnamefont{H.~D.}~\bibnamefont{Lutz}} \emph{et~al.},
  \bibinfo{journal}{J. Solid State Chem.} \textbf{\bibinfo{volume}{48}},
  \bibinfo{pages}{196} (\bibinfo{year}{1983}).





\end{thebibliography}
\end{document}